\documentclass{elsart}
\usepackage{graphicx}
\usepackage{amssymb}

\begin{document}

\begin{frontmatter}

\title{Convergence acceleration of series through a variational approach\thanksref{label1}}
\thanks[label1]{This work has been partially supported by CONACYT, grant 40633. The author wishes to thank R. S\'aenz for 
useful conversations and for reading the manuscript.}
\author{Paolo Amore}
\ead{paolo@ucol.mx}
\address{Facultad de Ciencias, Universidad de Colima,\\
Bernal D\'{i}az del Castillo 340, Colima, Colima,\\
Mexico.}

\title{Convergence acceleration of series through a variational approach}


\begin{abstract}
By means of a variational approach we find new series representations both for well known mathematical 
constants, such as $\pi$ and the Catalan constant, and for mathematical functions, such as the Riemann 
zeta function. The series that we have found are all exponentially convergent and provide 
quite useful analytical approximations. With limited effort our method can be applied to obtain 
similar exponentially convergent series for a large class of mathematical functions.
\end{abstract}


\end{frontmatter}

\section{Introduction}
\label{intro}

This paper deals with the problem of improving the convergence of a slowly convergent series where 
a large number of terms is needed to reach the desired accuracy. This is a challenging problem, 
which has been considered before (see, for example, \cite{FV96,Brad,CRZ:00}) and which has interesting 
applications in many physical problems. As a matter of fact it is well known that perturbation theory 
often gives series which converge very slowly or do not converge at all (one example is the perturbative 
series for the quantum anharmonic oscillator, which was originally studied in ~\cite{BW}).

In this paper we propose a new method to accelerate some class of mathematical series, which does not rely
on a perturbative approach, i.e. on an expansion in some small ¨natural¨ parameter (a parameter present in the 
original expression). The method works by introducing an artificial dependence in the formulas upon an 
arbitrary parameter, by identifying a new ``perturbation'' and by then devising an expansion which can be 
optimized to give faster rates of convergence. The details of how this works will be explained in depth in 
the next section. This procedure is well known in Physics and it has been exploited in the so-called 
``Linear Delta Expansion'' (LDE) and Variational Perturbation Theory (VPT) approaches~\cite{AFC90,Jo95,Fe00,Klei04}.

We will show in the following that such techniques can be used to obtain exponentially convergent series 
for some mathematical functions. A proof of convergence of the method is also provided. 
The ``flexibility'' and simplicity of the method that we propose suggests that application 
to wider classes of series could be found.

The paper is organized as follows: in section ~\ref{sec_mc} we first introduce the method and then use it to obtain 
accelerate series for $\pi$ and for the Catalan constant; in sections \ref{sec_zf} and \ref{sec_hf} we obtain a family of series 
representations for the Riemann and Hurwitz zeta functions which all converge in a certain domain of the arbitrary
parameter; finally in section \ref{concl} we draw our conclusions.

\section{Mathematical constants}
\label{sec_mc}

Many fundamental mathematical constants can be expressed as infinite sums. In many cases such series converge very slowly
and a huge number of terms has to be calculated before reaching the desired precision. Several examples of this 
are discussed for example in \cite{FV96}, where the authors consider a particular rearrangement of the series which 
transforms them into rapidly converging series. In the following we review two of the examples of \cite{FV96},
$\pi$ and the Catalan constant, and obtain new series representations for these constants which display a fast rate of convergence.

We first consider the series 
\begin{eqnarray}
S &=& 4 \sum_{n=1}^\infty \ \left[ \frac{1}{4 n -3} -\frac{1}{4 n-1} \right] \ , 
\label{eq1}
\end{eqnarray}
which converges very slowly to $\pi$ and it is known as Gregory's formula.  
The sum of the to the first $10^3$ terms yields an approximate value of $\pi$, 
which only has the first $3$ decimals correct.

Flajolet and Vardi ~\cite{FV96} have shown that it is possible to convert series such as the one in equation~(\ref{eq1}) into 
rapidly converging ones. Here we generalize the method of Flajolet and Vardi, introducing an arbitrary parameter in the series. 
Such parameter is then tuned to accelerate the convergence of the series itself by using the principle of minimal sensitivity 
(PMS)~\cite{Ste81}.

\begin{thm}
\label{thm1}
The series 
\begin{eqnarray}
S = 4 \sum_{m=1}^\infty \left(\frac{1}{1+\lambda}\right)^{m+1} \ \sum_{k=1}^m \left( \begin{array}{c}
m \\
k \\
\end{array} \right) \  \lambda^{m-k} \ \frac{3^k-1}{4^{k+1}} \ \zeta(k+1) \nonumber 
\end{eqnarray}
converges to $\pi$ for $\lambda > -1/2$, with $\lambda$ real.
\end{thm}

\begin{pf*}{Proof}

We write the series of equation~(\ref{eq1}) in the equivalent form
\begin{eqnarray}
S &=& \sum_{n=1}^\infty \frac{1}{n} \frac{1}{1+\lambda}  \left( \frac{1}{1- \frac{\frac{3}{4 n}+\lambda}{1+\lambda}} -
\frac{1}{1- \frac{\frac{1}{4 n}+\lambda}{1+\lambda}}\right)  \ ,
\label{eq2}
\end{eqnarray}
with an arbitrary parameter $\lambda \neq -1$.

Provided that $\left|\frac{\frac{3}{4 n}+\lambda}{1+\lambda}\right| < 1$ and 
$\left|\frac{\frac{1}{4 n}+\lambda}{1+\lambda}\right|<1$, for all $n \geq 1$, i.e. $\lambda > - 1/2$, we can expand equation~(\ref{eq2}) as
\begin{eqnarray}
S &=&  4 \sum_{n=1}^\infty \frac{1}{4 n}  \sum_{m=1}^\infty \left(\frac{1}{1+\lambda}\right)^{m+1} \
\sum_{k=1}^m \left( \begin{array}{c}
m \\
k \\
\end{array} \right) \  \lambda^{m-k} \ \left[ \left(\frac{3}{4 n}\right)^k - \left(\frac{1}{4 n}\right)^k \right]
\nonumber  \ .
\end{eqnarray}

As the series in $m$ and $n$ contain only positive terms, we can perform the series
over $n$ and obtain the result
\begin{eqnarray}
S &=&  \sum_{m=1}^\infty \left(\frac{1}{1+\lambda}\right)^{m+1} \ \sum_{k=1}^m \left( \begin{array}{c}
m \\
k \\
\end{array} \right) \  \lambda^{m-k} \ \frac{3^k-1}{4^{k}} \ \zeta(k+1) \ .
\label{eq3}
\end{eqnarray}

\end{pf*}

\begin{rem}
The series of equation~(\ref{eq3}) for $\lambda = 0$ coincides with the result of Flajolet and Vardi~\cite{FV96}:
\begin{eqnarray}
S^{(FV)} = \sum_{m=1}^\infty \frac{3^m-1}{4^m} \ \zeta(m+1) 
\label{eq4}
\end{eqnarray}
and converges geometrically to $\pi$.

\end{rem}

Notice that equation~(\ref{eq3}) defines a family of series converging to $\pi$ as long as $\lambda > - 1/2$. 
Clearly the dependence upon $\lambda$ in  equation~(\ref{eq3}) is artificial and shows up only when a finite number 
of terms is considered. If we set $S_N(\lambda)$ to be the partial series of  equation~(\ref{eq3})
the dependence upon $\lambda$ in $S_N(\lambda)$  then  disappears in the limit $N \rightarrow \infty$. 

For fixed $N$ we evaluate the partial sum at the points where \[dS_N(\lambda)/d\lambda=0\], since 
there the expression is less sensitive to changes of the arbitrary parameter $\lambda$, a property which shares with the 
full series (\ref{eq3}). This is called  Principle of Minimal Sensitivity (PMS)~\cite{Ste81} and provides an equation
which, once solved at a given order, provides an {\sl optimal} value of $\lambda$ for a fixed partial sum $S_N(\lambda)$.

For $S_2(\lambda)$ we obtain 
\begin{eqnarray}
\lambda^{(1)} = - \frac{3}{\pi^2} \ \zeta(3) \approx - 0.365381 > - 1/2 \ .
\label{eq5}
\end{eqnarray}

\begin{rem}
The series of equation~(\ref{eq3}) converges geometrically to $\pi$. We can estimate the rate of convergence by approximating 
the $m$th term in the series with
\begin{eqnarray}
s_m &=&  \left(\frac{1}{1+\lambda}\right)^{m+1} \ \sum_{k=1}^m \left( \begin{array}{c}
m \\
k \\
\end{array} \right) \  \lambda^{m-k} \ \frac{3^k-1}{4^{k}} \nonumber \\
&=& \frac{1}{(1+\lambda)^{m+1}} \ 
\left[ (\lambda+3/4)^m -  (\lambda+1/4)^m \right] \approx m  \left(\frac{\lambda+3/4}{1+\lambda} \right)^m \ ,
\end{eqnarray}

Using the PMS value of equation (\ref{eq5}) we obtain  $s_m \approx 1.65^{-m}$. This improves the rate $s_m \approx 1.33^{-m}$ 
of the series of Flajolet and Vardi (\ref{eq4}).

\end{rem}

\begin{figure}
\begin{center}
\includegraphics[width=10cm]{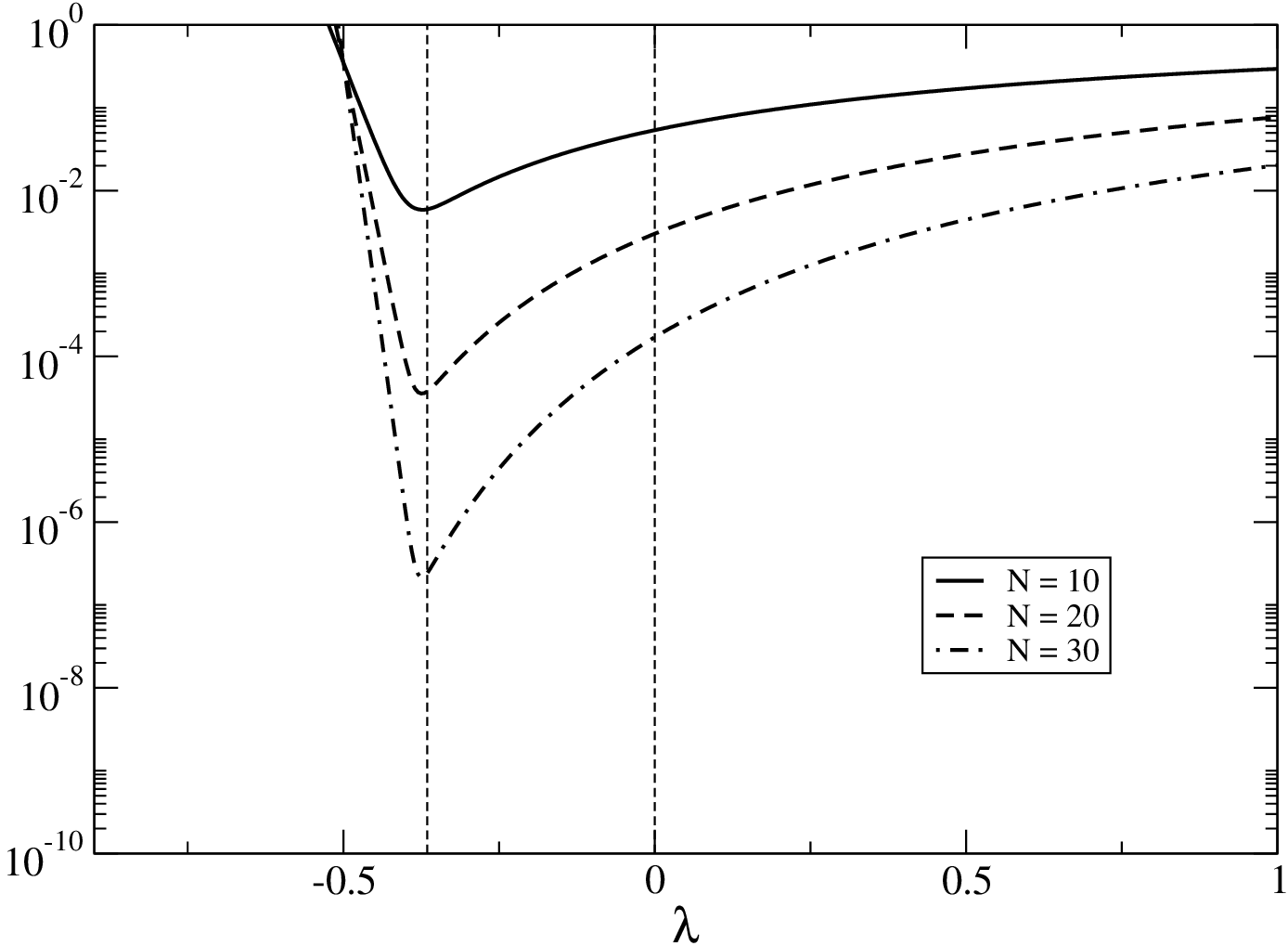}
\caption{The error obtained using the partial sum of Equation~(\ref{eq3})
over the first $10$, $20$ and $30$ terms respectively as a function of $\lambda$.\label{fig1}}
\end{center}
\end{figure}

\begin{figure}
\begin{center}
\includegraphics[width=10cm]{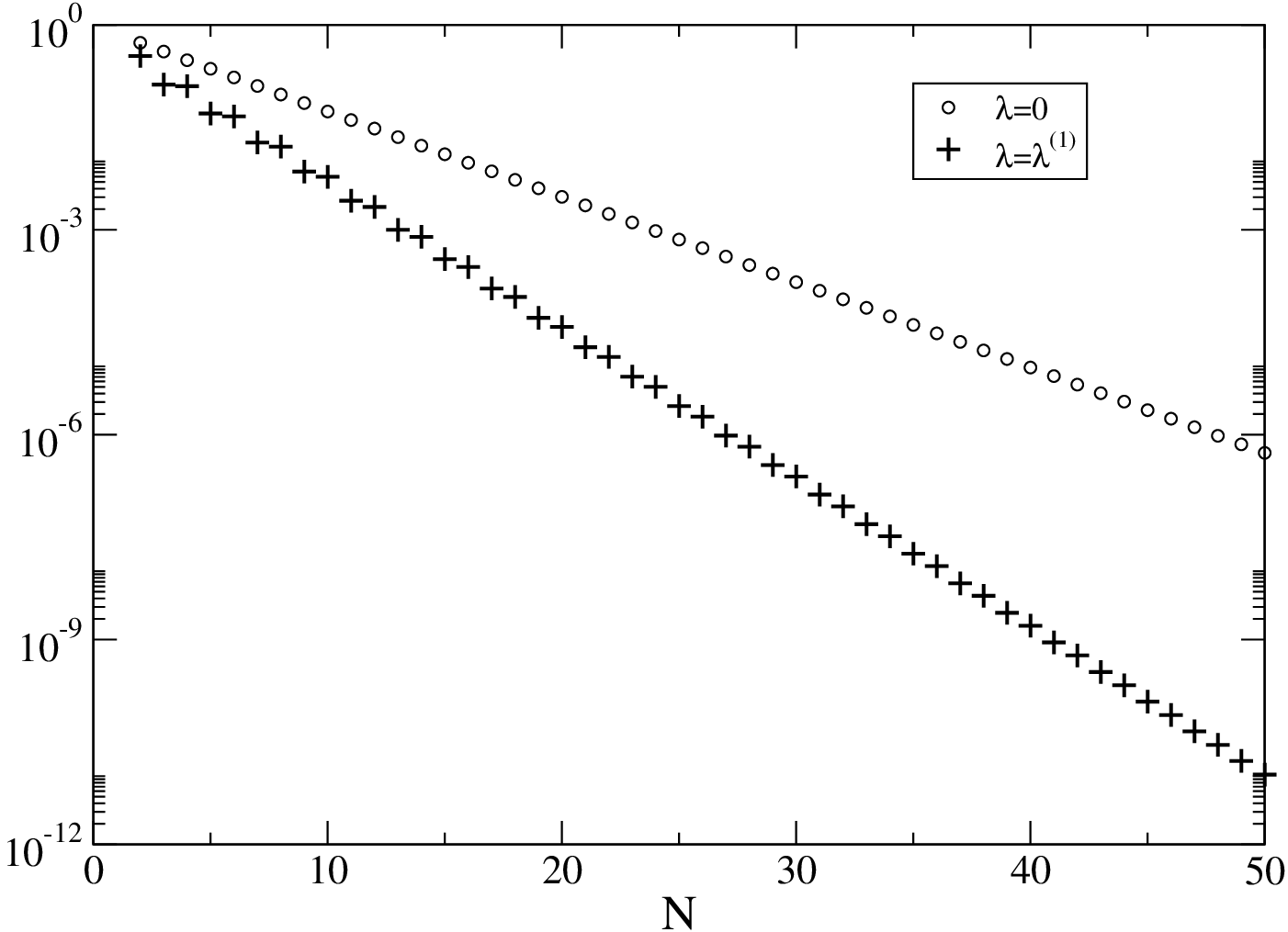}
\caption{The error obtained using the partial sum of Equation~(\ref{eq3}) with $\lambda=0$ and
$\lambda=\lambda^{(1)}$ as a function of the number of terms in the sum.\label{fig2}}
\end{center}
\end{figure}

In Figure~\ref{fig1} we display the partial sums of equation~(\ref{eq3}) with $10$, $20$ and $30$ terms as a function of 
$\lambda$: the locations of $\lambda^{(1)}$  and $\lambda =0$ are marked with vertical lines. 
It turns out that $\lambda^{(1)}$ is an excellent approximation to the exact minimum of the partial sum even for large 
values of terms. 

In Figure~\ref{fig2} we plot the error obtained by using  equation~(\ref{eq3}) with $\lambda$ given by equation~(\ref{eq5}) 
and by using the formula of Flajolet and Vardi, equation~(\ref{eq4}). Our series converges exponentially
more rapidly than equation~(\ref{eq4}).

We now consider another series, which was also considered in \cite{FV96}. The series 
\begin{eqnarray}
S = \sum_{n=1}^\infty \left[ \frac{1}{(4 n-3)^2} - \frac{1}{(4 n-1)^2} \right]
\label{eq6}
\end{eqnarray}
is known to slowly converge to the Catalan constant, $G \approx 0.9159656$.

\begin{thm} 
\label{thm2}
The series defined as 
\begin{eqnarray}
S &=&  \sum_{m=1}^\infty \left(\frac{1}{1+\lambda}\right)^{m+1} \ \sum_{k=1}^m \left( \begin{array}{c}
m \\
k \\
\end{array} \right) \  \lambda^{m-k} \ k \ \frac{3^{k-1}-1}{4^{k+1}} \ \zeta(k+1) \ ,
\label{eq9}
\end{eqnarray}
converges to the Catalan constant for any $\lambda > -1/2$, with $\lambda$ real.
\end{thm}

\begin{pf*}{Proof}
We can rewrite equation~(\ref{eq6}) as
\begin{eqnarray}
S &=& \lim_{a\rightarrow 0} \frac{d}{da} \tilde{S}(a) 
\label{eq7}
\end{eqnarray}
where
\begin{eqnarray}
\tilde{S} (a) &\equiv&  \sum_{n=1}^\infty \left[ \frac{1}{(4 n-3-a)} - \frac{1}{(4 n-1-a)} \right]  \ .
\nonumber 
\end{eqnarray}

Note that this series converges uniformly in $a$ so we can differentiate term by term. We can apply Theorem
\ref{thm1} to $\tilde{S}(a)$  and obtain
\begin{eqnarray}
\tilde{S}(a) &=& \sum_{m=1}^\infty \ \sum_{k=1}^m \left( \begin{array}{c}
m \\
k \\
\end{array} \right) \ \frac{\lambda^{m-k}}{(1+\lambda)^{m+1}}  \ \frac{(3+a)^k-(1+a)^k}{4^{k+1}} \ \zeta(k+1) \ .
\label{eq8a}
\end{eqnarray}

As the series (\ref{eq8a}) converges uniformly in $a$, the proof is complete once the limit (\ref{eq7}) is evaluated.
\end{pf*}

Notice that the series of Theorem \ref{thm2} reduces to the formula given in \cite{FV96} for $\lambda=0$:
\begin{eqnarray}
S^{(FV)} &=&  \sum_{m=1}^\infty m \ \frac{3^{m-1}-1}{4^{m+1}} \ \zeta(m+1) \ .
\label{eq9a}
\end{eqnarray}

\begin{rem} The series defined as 
\begin{eqnarray}
S &=&  \frac{1}{3}  \sum_{m=1}^\infty \ \sum_{k=1}^m \frac{1}{2^{3+2 k}} \ \frac{\lambda_0^{m-k}}{(1+\lambda_0)^{m+2}} 
\left( \begin{array}{c}
m \\
k \\
\end{array} \right) \zeta(k+1) \nonumber \\
&\times& \left[ - \left( \left( 3 + 3^k \right) \ k \ \left( 1 + \lambda_0 \right)  \right) 
- 3 \ \left( -1 + 3^k \right) \ \left( \lambda_0 - m \right)  \right]
\label{eq11}
\end{eqnarray}
converges to the Catalan constant, where $\lambda_0 \equiv -  \frac{3}{\pi^2} \ \zeta(3)$. 

This follows from equation~(\ref{eq8a}) for $\tilde{S}(a)$ applying  the PMS \[\frac{d\tilde{S}(a)}{d\lambda} = 0 . \] 
The optimal value of $\lambda$  taking the first two terms in the series
\begin{eqnarray}
\lambda^{(1)} = -  \frac{3}{\pi^2} \ \zeta(3) \left(1 + \frac{a}{2} \right) \ ,
\label{eq10}
\end{eqnarray}
and equation (\ref{eq11}) is obtained.

\end{rem}

\begin{figure}
\begin{center}
\includegraphics[width=10cm]{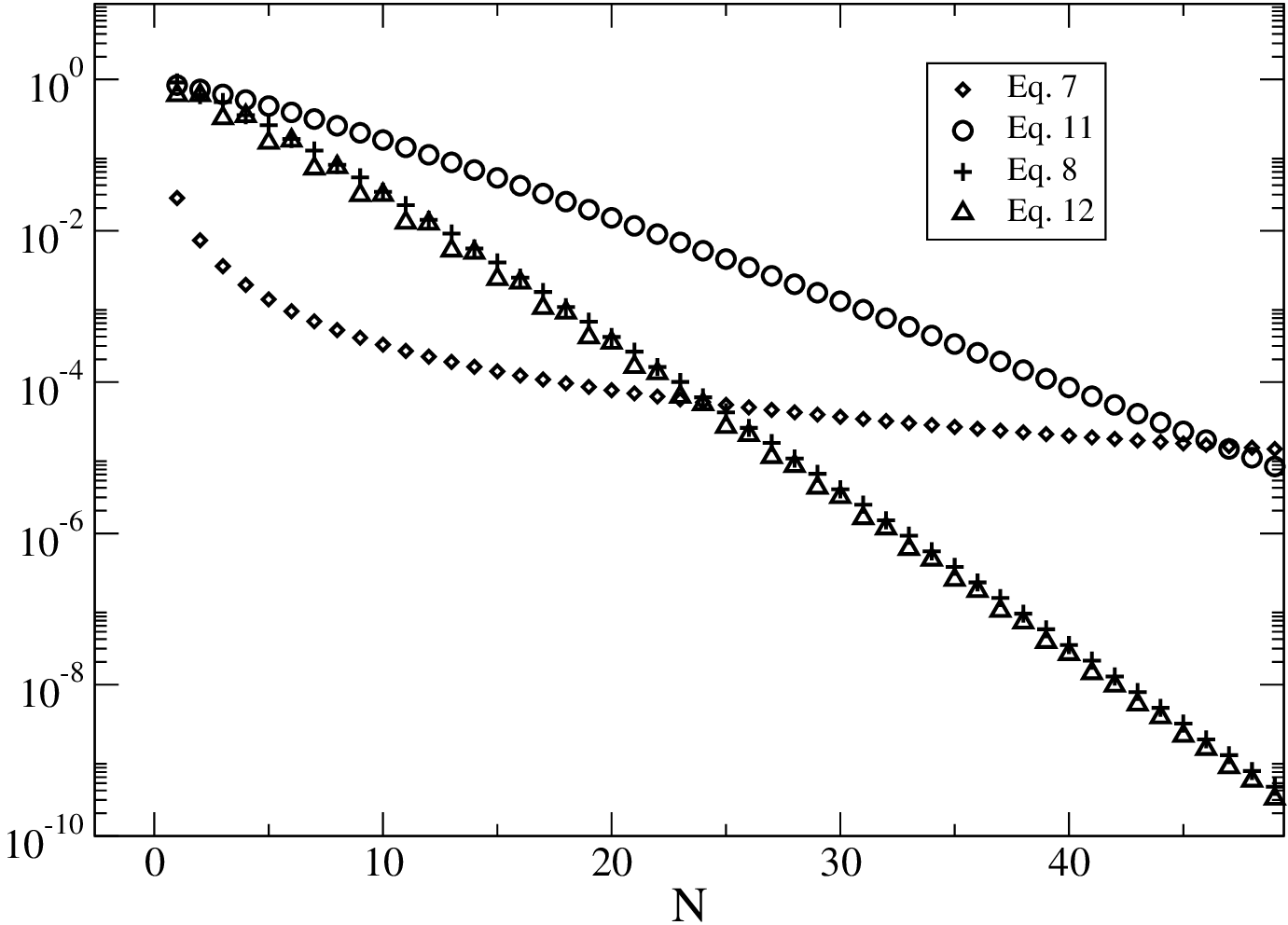}
\caption{$|S^{(N)} - G|$ as a function of the number of terms in the sum using equation~(\ref{eq6}) (diamonds), 
equation~(\ref{eq9a}) (circles), equation~(\ref{eq9}) (pluses) and equation~(\ref{eq11}) (triangles).\label{fig3}}
\end{center}
\end{figure}

In Figure~\ref{fig3} we compare the different approximations, showing that equation~(\ref{eq9}) (with $\lambda = \lambda^{(1)}$)
and equation~(\ref{eq11}) have a greater rate of convergence then the corresponding equation in \cite{FV96}. 
Equation~(\ref{eq11}) provides a  slightly better approximation.

\begin{rem}
The accelerated series (\ref{eq9}) and (\ref{eq11}) converge geometrically to the Catalan constant. For example the $m$th 
term of the series  (\ref{eq9}) behaves as
\begin{eqnarray}
s_m \approx  \frac{m}{16 \ (1+\lambda)^{m+1}} \ \left[\left(\lambda+3/4 \right)^{m-1} -\left(\lambda+1/4 \right)^{m-1}  \right]
\approx  m  \left(\frac{\lambda+3/4}{1+\lambda} \right)^m \ .
\end{eqnarray}

Taking $\lambda=\lambda_0$ we have  $s_m \approx m \ 1.65^{-m}$.

\end{rem}

It is clear that these  results can be generalized to sums of the form
\begin{eqnarray}
S_n = \sum_{n=1}^\infty \left[ \frac{1}{(4 n-3)^n} - \frac{1}{(4 n-1)^n} \right] \ .
\end{eqnarray}

\section{The Riemann zeta function}
\label{sec_zf}

In this section we apply the same strategy outlined above to the calculation of the Riemann zeta function~\cite{BBC00},
and prove the following theorem
\begin{thm}
A convergent series for the Riemann zeta function, which is valid for $\Re (s)>0$, with the exclusion of $s=1$, is
\begin{eqnarray}
\zeta(s) &=&  \frac{1}{1-2^{1-s}}  \sum_{k=0}^\infty \sum_{j=0}^k \ \left( \begin{array}{c}
k \\
j \\
\end{array} \right) \ \frac{\lambda^{k-j}  }{(1+\lambda)^{k+1}}  \frac{(-1)^{j} }{(1+j)^s} \ ,
\label{s4_3}
\end{eqnarray}
for $\lambda >0$.
\end{thm}

\begin{pf*}{Proof}
We use the integral representation
\begin{eqnarray}
\zeta(s) &=&  \frac{1}{1-2^{1-s}} \frac{1}{\Gamma(s)} \int_0^1 \frac{\log^{s-1} \frac{1}{x}}{1+x} dx \ ,
\label{s4_1}
\end{eqnarray}
valid for $\Re(s) >0$, and write it as
\begin{eqnarray}
\zeta(s) &=& \frac{1}{1-2^{1-s}} \frac{1}{\Gamma(s)} \int_0^1 \frac{1}{1+\lambda} \ 
\frac{\log^{s-1} \frac{1}{x}}{1+\frac{x-\lambda}{1+\lambda}} dx  ,
\label{s4_2}
\end{eqnarray}
where $\lambda$ is an arbitrary parameter introduced by hand.
The condition $\left|\frac{x-\lambda}{1+\lambda}\right| < 1$ is fullfilled uniformly for all $x \in [0,1]$ 
provided that $\lambda >0$; in this case one can expand the denominator in powers of 
$\left(\frac{x-\lambda}{1+\lambda}\right)$ and obtain
\begin{eqnarray}
\zeta(s) &=&  \frac{1}{1-2^{1-s}} \frac{1}{\Gamma(s)} \sum_{k=0}^\infty \frac{1}{(1+\lambda)^{k+1}} 
\int_0^1  \left(-x+\lambda\right)^k \log^{s-1} \frac{1}{x}  \ dx \nonumber \\
&=&  \frac{1}{1-2^{1-s}}  \sum_{k=0}^\infty \sum_{j=0}^k \  \left( \begin{array}{c}
k \\
j \\
\end{array} \right) \ \frac{\lambda^{k-j}  }{(1+\lambda)^{k+1}}  \frac{(-1)^{j} }{(1+j)^s} \ ,
\end{eqnarray}
which completes our proof.
\end{pf*}

\begin{rem}
The series (\ref{s4_3}) for $\lambda=1$,
\begin{eqnarray}
\zeta(s) &=&  \frac{1}{1-2^{1-s}}  \sum_{k=0}^\infty \frac{1}{2^{k+1}}  \ \sum_{j=0}^k \ \left( \begin{array}{c}
k \\
j \\
\end{array} \right) \ \frac{(-1)^{j} }{(1+j)^s} \ ,
\label{khseq}
\end{eqnarray} 
has been first conjectured by Knopp around 1930~\cite{Kno30}, and later
proved by Hasse~\cite{Has30} and rediscovered more recently by Sondow~\cite{Son94}.
\end{rem}

Although $\lambda$ appears explicitly in the series (\ref{s4_3}), the series itself  does not depend upon $\lambda$, 
as long as $\lambda >0$. However the partial sums $\zeta^{(K)}(\lambda,s)$, defined as
\begin{eqnarray}
\zeta^{(K)}(\lambda,s) \equiv \frac{1}{1-2^{1-s}}  \sum_{k=0}^{K} \sum_{j=0}^k \  \left( \begin{array}{c}
k \\
j \\
\end{array} \right) \ \frac{\lambda^{k-j}  }{(1+\lambda)^{k+1}}  \frac{(-1)^{j} }{(1+j)^s}  \ ,
\end{eqnarray}
must show a dependence upon $\lambda$. However such dependence may be minimized by applying the PMS, i.e.
\begin{eqnarray}
\frac{d}{d\lambda} \zeta^{(K)}(\lambda,s)= 0 \ .
\end{eqnarray}

To lowest order, which corresponds to sum up to $K=1$, one has that $\lambda_{PMS}^{(1)} = 2^{-s}$ and the 
corresponding formula is found to be:
\begin{eqnarray}
\zeta(s) &=&   \frac{1}{1-2^{1-s}} \ \sum_{k=0}^\infty \frac{1}{(1+2^{-s})^{k+1}} \sum_{j=0}^k \ 
 \left( \begin{array}{c}
k \\
j \\
\end{array} \right) \  2^{-s (k-j)}  \frac{(-1)^{j} }{(1+j)^s}  .
\label{s4_4}
\end{eqnarray}

\begin{figure}
\begin{center}
\includegraphics[width=10cm]{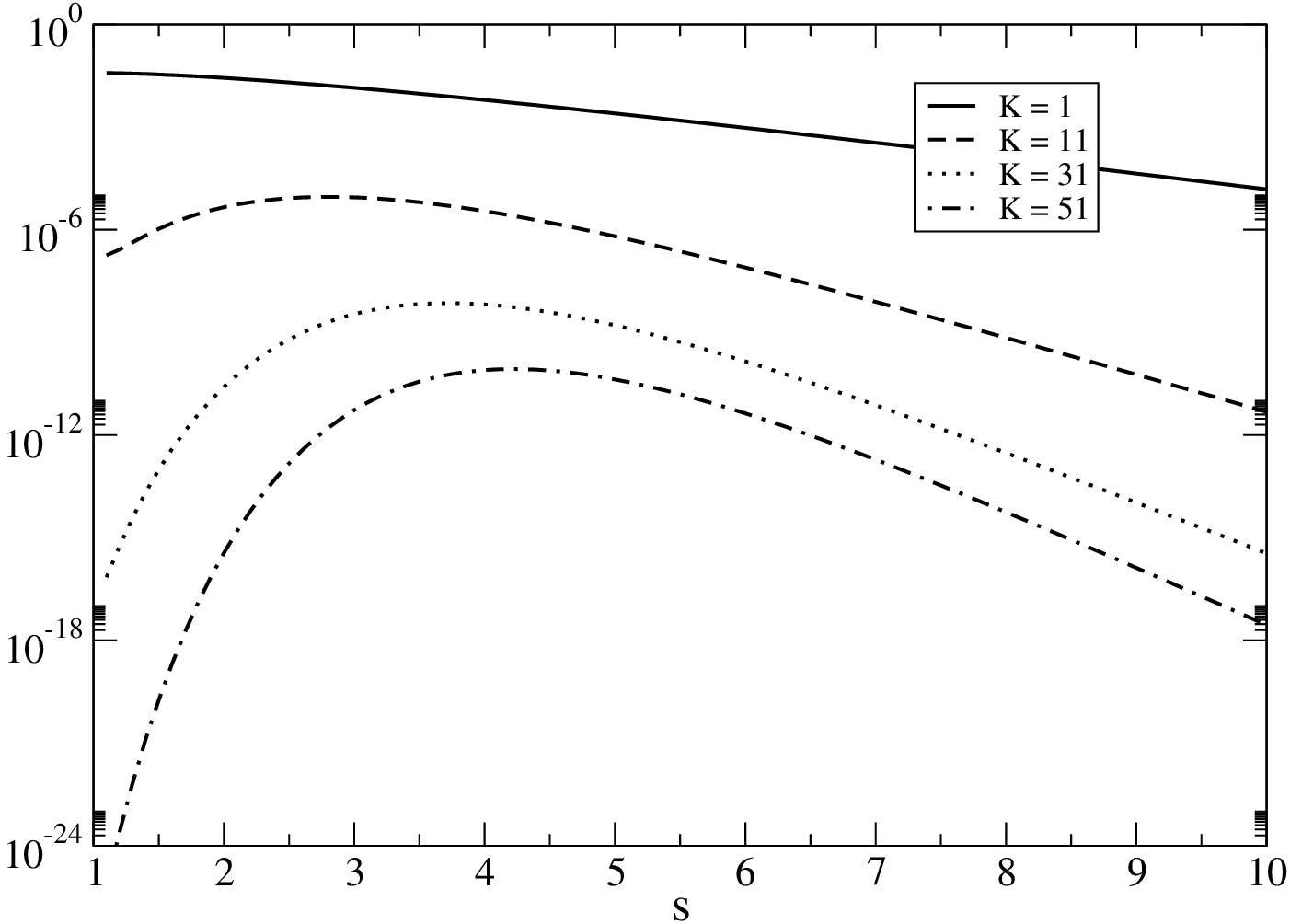}
\caption{The error $\left| (\zeta^{(K)}(s)-\zeta(s))/\zeta(s) \right|$ obtained by using equation~(\ref{s4_4}) 
to different orders. \label{fig4}}
\end{center}
\end{figure}

Equation~(\ref{s4_4}) is an {\sl exact} series representation of the Riemann zeta function:
this simple formula yields an excellent approximation to the zeta function as it can be appreciated by looking
at Figure~\ref{fig4}, where we plot the error $\left| (\zeta^{(K)}(s)-\zeta(s))/\zeta(s) \right|$ obtained by 
using equation~(\ref{s4_4}) to different orders, for $1<s \leq 10$. It is remarkable that this simple analytical formula works
quite well even in the proximity of $s=1$, where the $\zeta$ function diverges.

\begin{figure}
\begin{center}
\includegraphics[width=10cm]{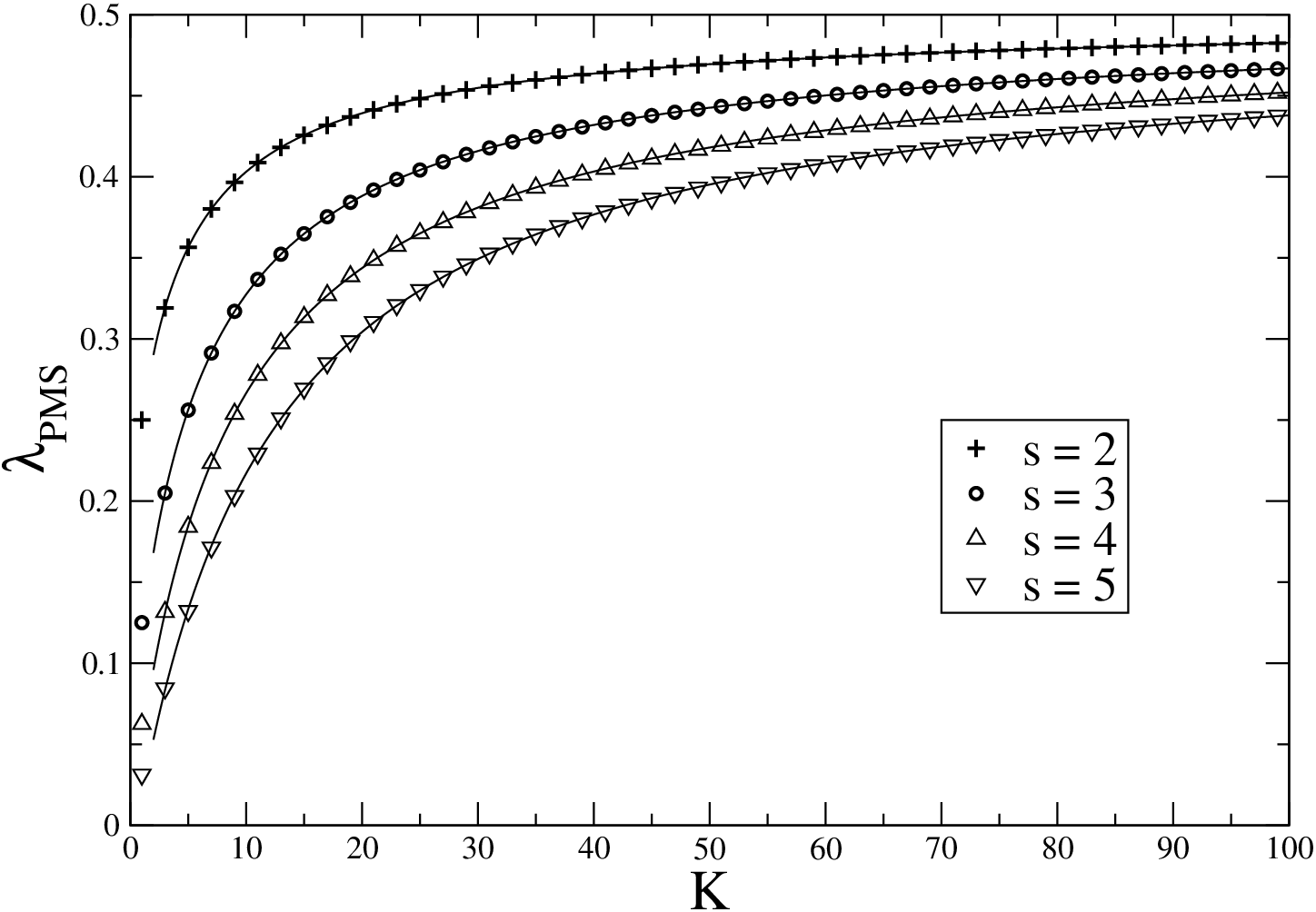}
\caption{The optimal parameter $\lambda_{PMS}$ for $s=2,3,4,5$ calculated to different orders.
\label{fig4a}}
\end{center}
\end{figure}

The rate of convergence of the series is greatly improved by applying the PMS to higher 
orders\footnote{A real solution is found only for odd values of $K$.}. Although it is 
possible to find the analytical solution to the PMS equation only to low orders, we have 
calculated $\lambda$ numerically in Figure~\ref{fig4a} for $s=2,3,4,5$. 
For example, to order $K=101$ we find that $\lambda_{PMS}^{(101)}(2) = 0.482$, 
$\lambda_{PMS}^{(101)}(3) = 0.467$, $\lambda_{PMS}^{(101)}(4) = 0.452$ and 
$\lambda_{PMS}^{(101)}(5) = 0.439$.

\begin{rem}
The thin solid lines correspond to the best fit of $\lambda$ with the function
\begin{eqnarray}
\lambda_{FIT}(K) = \kappa_1 + \frac{\kappa_2 \ K \ \log K}{\kappa_3 K + (\kappa_4+K) \ \log K} \nonumber \ ,
\end{eqnarray}
for $K$ up to $101$. The coefficients $\kappa_i$ for $s=2,3,4,5$ can be found in Table \ref{table1}.
\end{rem}

\begin{table}
\begin{center}
\begin{tabular}{|c|cccc|}
\hline
$s$ & $\kappa_1$ & $\kappa_2$ & $\kappa_3$ & $\kappa_4$ \\
\hline
$2$ & $0.201$ &  $0.318$ & $0.416$ & $3.9396$ \\
$3$ &  $0.0655$ &  $0.461$ & $0.415$ & $5.791$ \\
$4$ &  $0.0035$ &  $0.514$ & $0.288$ & $8.283$ \\
$5$ &  $-0.0245$ &  $0.515$ & $0.0031$ & $11.31$ \\
\hline
\end{tabular}
\end{center}
\bigskip
\caption{Coefficients of the best fit of the $\lambda_{PMS}$ as a function of $K$.\label{table1}}
\end{table}

\begin{figure} 
\begin{center}
\includegraphics[width=10cm]{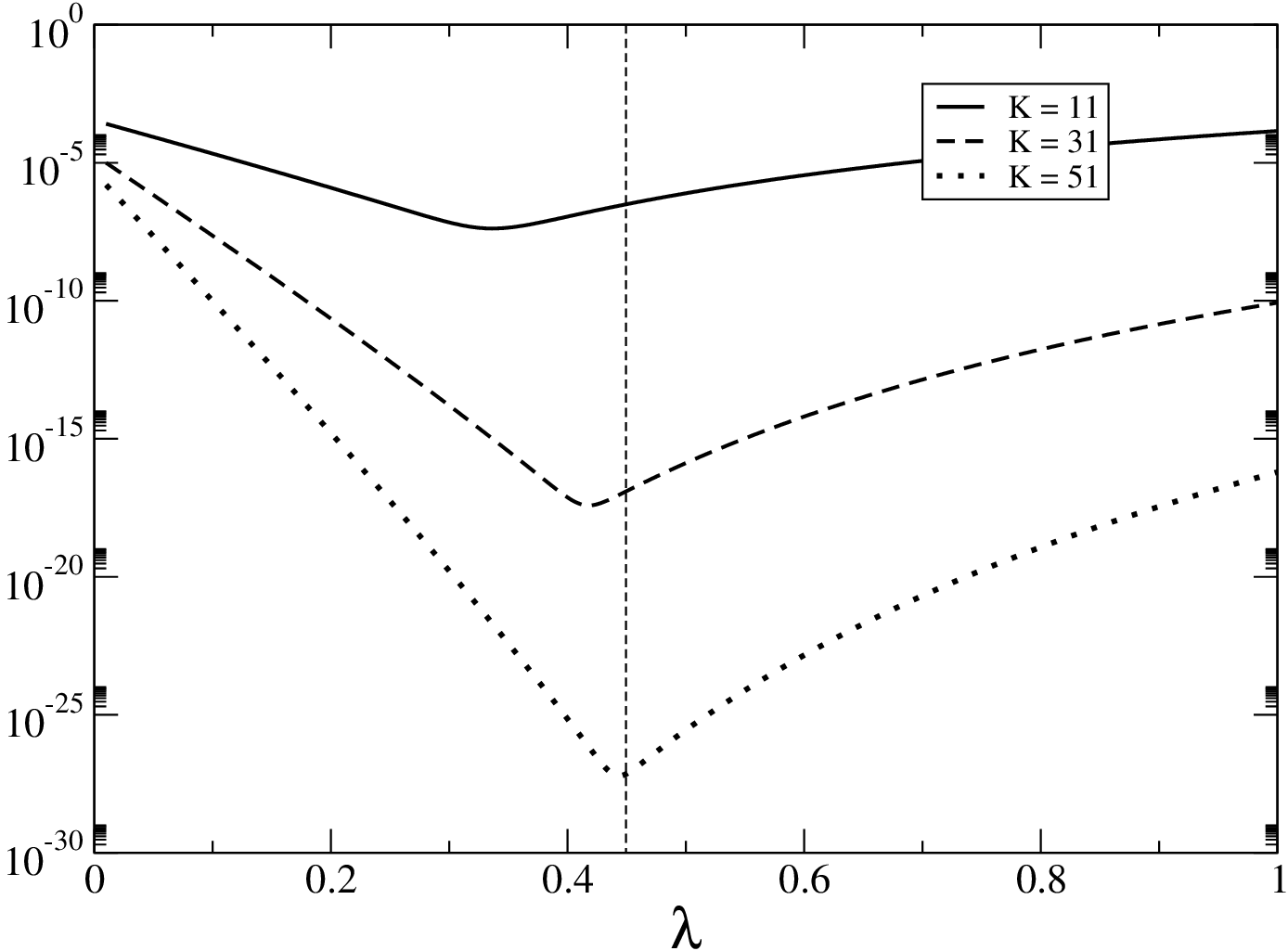}
\caption{Dependence upon the variational parameter of 
$\left| (\zeta^{(K)}(s)-\zeta(s))/\zeta(s) \right|$, with $K=11,31,51$.\label{fig5}}
\end{center}
\end{figure}

\begin{figure}
\begin{center}
\includegraphics[width=10cm]{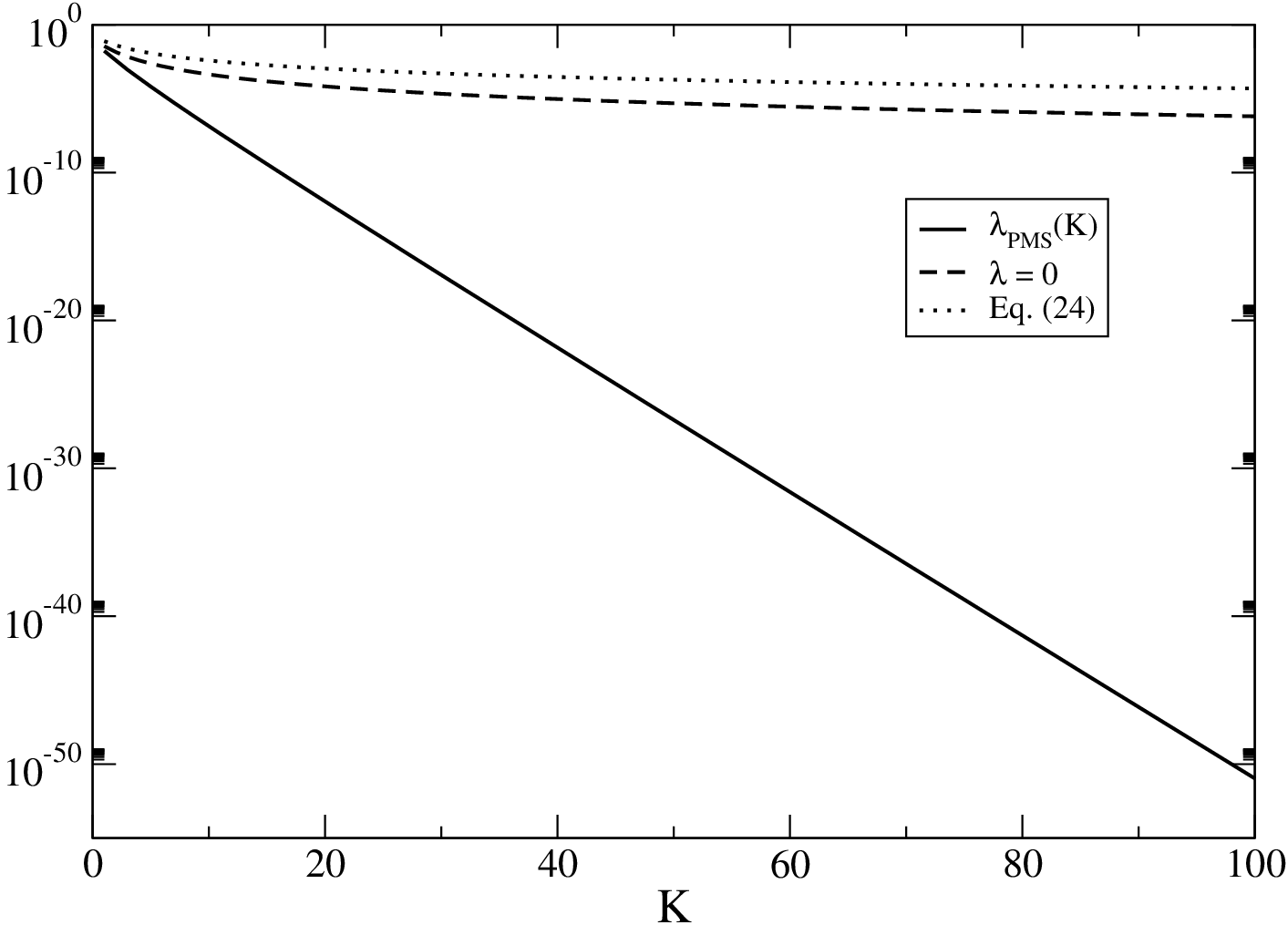}
\caption{Difference  $|\zeta^{(K)}(3)-\zeta(3)|$  as a function of the number of terms in the sum.\label{fig6}}
\end{center}
\end{figure}

In Figure~\ref{fig5} we display the dependence upon $\lambda$ of the partial sums over the first $K=11,31,51$ terms
in the case of $\zeta(3)$. The vertical line corresponds to the location of $\lambda_{PMS}^{(101)}$.
In Figure~\ref{fig6} we plot the difference $|\zeta^{(K)}(3)-\zeta(3)|$ using equation~(\ref{s4_3}) with 
$\lambda = \lambda_{PMS}^{(K)}$ (solid line), $\lambda = 0$ (dashed line) and the series representation
\begin{eqnarray}
\zeta(3) = \sum_{n=0}^\infty \frac{1}{(n+1)^3} 
\label{s4_5}
\end{eqnarray}
which corresponds to the dotted line in the plot. This last series converges quite slowly and a huge number
of terms (of the order of $10^{25}$) is needed to obtain the same accuracy that our series  
with $\lambda_{PMS}^{(K)}$ reaches with just $10^2$ terms.

\begin{rem}
We now define
\begin{eqnarray}
c_K(\lambda,s) &\equiv& \sum_{j=0}^K \ \left( \begin{array}{c}
K \\
j \\
\end{array} \right) \ \frac{\lambda^{K-j}  }{(1+\lambda)^{K+1}}  \frac{(-1)^{j} }{(1+j)^s} \ ,
\end{eqnarray}
which we can write as
\begin{eqnarray}
c_K(\lambda,s) &=& \sum_{j=0}^K \ \left( \begin{array}{c}
K \\
j \\
\end{array} \right) \ \frac{\lambda^{K-j}  }{(1+\lambda)^{K+1}}  \frac{(-1)^{j} }{\Gamma(s)} \  \int_0^\infty \ e^{-(1+j) t} \ t^{s-1} \ dt \nonumber \\
&=&  \frac{1}{\Gamma(s)} \  \int_0^\infty \ e^{- t} \ t^{s-1}  \ \frac{(\lambda-e^{-t})^K}{(1+\lambda)^{K+1}} \ dt \ .
\end{eqnarray}

For $\lambda>1$ we have the inequality
\begin{eqnarray}
|c_K(\lambda,s)| &\leq&  \frac{1}{\Gamma(s)} \  \int_0^\infty \ e^{- t} \ t^{s-1}  \ \frac{\lambda^K}{(1+\lambda)^{K+1}} \ dt = 
\frac{\lambda^K}{(1+\lambda)^{K+1}} \ .
\end{eqnarray}

For $0 < \lambda <1$ we can split the integral in the two regions $0 < t < \log\frac{1}{\lambda}$ and $t >\log\frac{1}{\lambda}$
and obtain the inequality
\begin{eqnarray}
|c_K(\lambda,s)| &\leq&  \frac{1}{\Gamma(s)} \  \int_0^{\log\frac{1}{\lambda}} \ e^{- t} \ t^{s-1}  \ 
\frac{(\lambda-e^{-t})^K}{(1+\lambda)^{K+1}} \ dt +  
\frac{1}{\Gamma(s)} \  \int_{\log\frac{1}{\lambda}}^\infty \ \frac{e^{- t (K+1)} \ t^{s-1}  }{(1+\lambda)^{K+1}} 
\ dt \nonumber \\
&\leq& \frac{\lambda^K}{(1+\lambda)^{K+1}} \left(1-\frac{\Gamma(s,\log\frac{1}{\lambda})}{\Gamma(s)}\right) \nonumber \\ 
&+& \frac{1}{(1+\lambda)^{K+1}} \ \frac{\Gamma(s, (K+1) \log\frac{1}{\lambda})}{\Gamma(s)} \frac{1}{(K+1)^s} \ ,
\end{eqnarray}
where $\Gamma(a,x)$ is the incomplete gamma function. Using the asymptotic behaviour of $\Gamma(a,x)$ (see 6.5.32 of \cite{AbS}),
\[
\Gamma(a,x) \approx x^{a-1} e^{-x}
\]
we can estimate the rate of convergence to be
\begin{eqnarray}
|c_K(\lambda,s)| &\lessapprox&   \frac{\lambda^K}{(1+\lambda)^{K+1}} \left[1 - \frac{\lambda}{\Gamma(s)} \ \left( \log\frac{1}{\lambda}\right)^{s-1} 
\left(1-  \frac{1}{1+K} \right) \right] \ ,
\end{eqnarray}
which is essentially geometric.

Notice that the $K$th term of series of equation (\ref{khseq})  decays with a much slower rate, given by
\begin{eqnarray}
c_K^{(KHS)} &\approx& 2^{-K} \ .
\end{eqnarray}

\end{rem}






If $s$ is on the critical line, i.e. $s=1/2+ i \tau$, we have
\begin{eqnarray}
\zeta\left(\frac{1}{2} + i \tau\right) &=& \frac{(2)^{-\frac{1}{2} + i \tau}}{2^{-\frac{1}{2} + i \tau}-1}  
\sum_{k=0}^\infty \sum_{j=0}^k \ 
  \left( \begin{array}{c}
k \\
j \\
\end{array} \right) \ \frac{\lambda^{k-j}}{(1+\lambda)^{k+1}} \ \frac{(-1)^{j}}{(1+j)^{\frac{1}{2} + i \tau}} 
\label{eq_complx}
\end{eqnarray}

\begin{figure}
\begin{center}
\includegraphics[width=10cm]{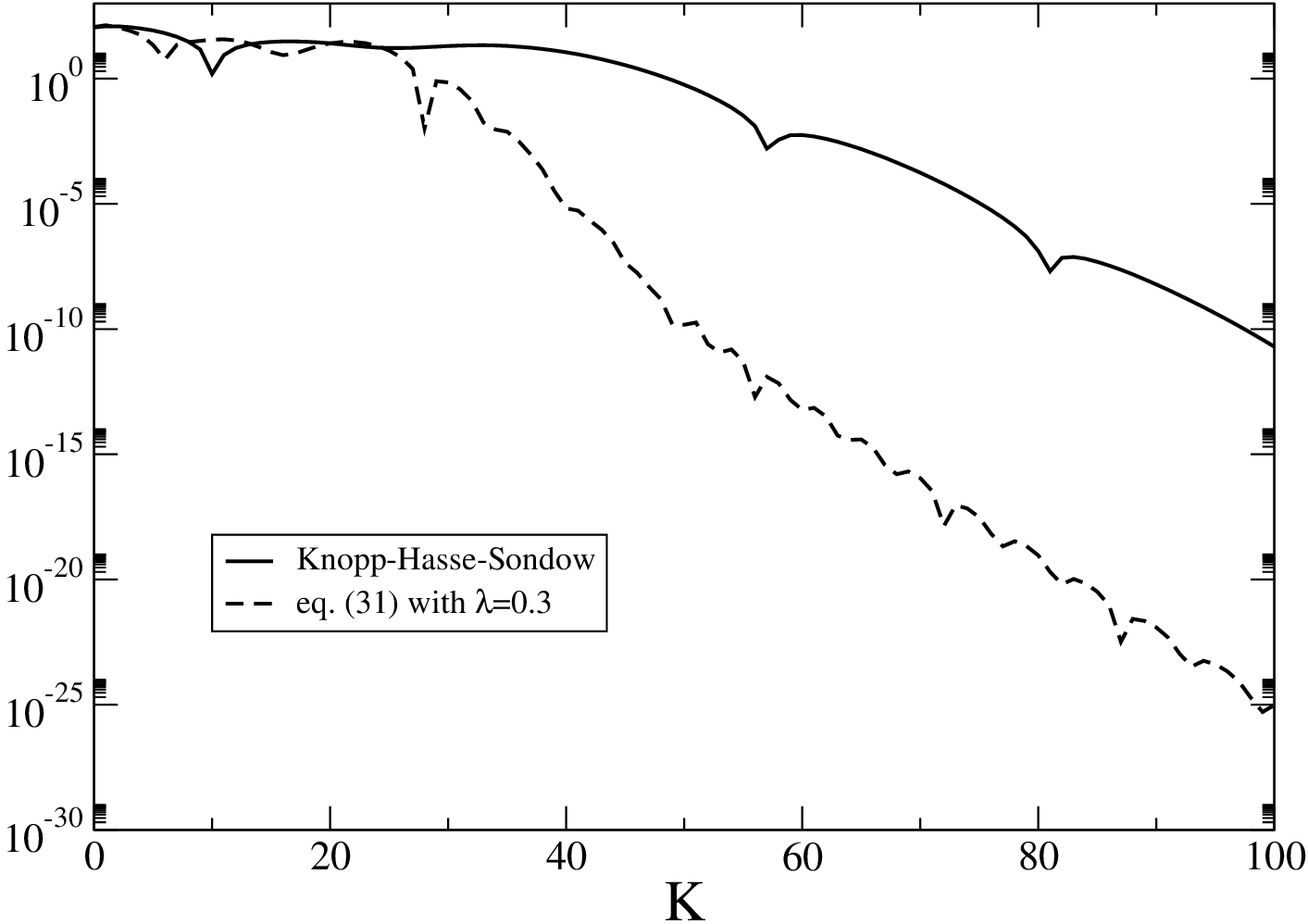}
\caption{$\Xi \equiv \Re \left[\frac{\zeta^{(K)}\left(\frac{1}{2} + i \tau\right) -
\zeta\left(\frac{1}{2} + i \tau\right)}{\zeta\left(\frac{1}{2} + i \tau\right)}\right] \times 100$, as a 
function of the number of terms considered in the sum of equation~(\ref{eq_complx}). The dashed curve is
obtained using $\lambda = 0.3$.
\label{figcomplx}}
\end{center}
\end{figure}

In Figure~\ref{figcomplx} we have plotted the error (in percent) over the real part of the zeta function, i.e.
$\Xi \equiv \Re \left[\frac{\zeta^{(K)}\left(\frac{1}{2} + i \tau\right) -\zeta\left(\frac{1}{2} + i \tau\right)}{\zeta\left(\frac{1}{2} + 
i \tau\right)}\right] \times 100$, as a function of the number of terms considered in the sum of equation~(\ref{eq_complx}). We use $\tau = 50$. 
The solid line corresponds to equation (\ref{khseq}), whereas the dashed line corresponds to using our formula, 
equation~(\ref{eq_complx}), with $\lambda = 0.3$. 

\section{The generalized Hurwitz zeta function}
\label{sec_hf}

We now turn our attention to the generalized Hurwitz zeta function given by
\begin{eqnarray}
\overline{\zeta}(s, u,\xi) &=& \sum_{n=0}^\infty \frac{1}{\left(n^u + \xi \right)^s} \ ,
\label{s3_1a}
\end{eqnarray}
which includes as special cases both the Riemann ($\overline{\zeta}(s,1,1)$) and the Hurwitz ($\overline{\zeta}(s,1,\xi)$) 
zeta functions.

We prove the following theorem:
\begin{thm}
Let $s$ and $u$ be real numbers such that $s u >1$; then 
\begin{eqnarray}
\overline{\zeta}(u,s,\xi)  &=& \frac{1}{\xi^s} + 
\sum_{k=0}^\infty \frac{\Gamma(k+s)}{\Gamma(s) } \ \Psi_{k}(\lambda,u,s,\xi) 
\label{eq_2}
\end{eqnarray}
where
\begin{eqnarray}
\Psi_{k}(\lambda,u,s,\xi) &\equiv&  \sum_{j=0}^k \ \frac{(-\xi)^j}{j! (k-j)!} 
\frac{\lambda^{2 (k-j)} }{(1+\lambda^2)^{s+k}}  \zeta(us+uj) 
\label{eq_3}
\end{eqnarray}
and $\lambda^2>\frac{\xi-1}{2}$ ($\xi > 0$).
\end{thm}

\begin{pf*}{Proof}

For $us>1$ the series $\overline{\zeta}(u,s,\xi)$ converges; we use the identity
\begin{eqnarray}
\overline{\zeta}(u,s,\xi) 
= \frac{1}{\xi^s} + \sum_{n=1}^\infty \frac{1}{n^{s u}} \  
\frac{1}{\left(1+\lambda^2\right)^s} \ \frac{1}{\left(1+ \Delta(n)\right)^s} \ ,
\label{eq_4}
\end{eqnarray}
where
\begin{eqnarray}
\Delta(n) &\equiv& \frac{\xi/n^u-\lambda^2}{1+\lambda^2} \ .
\label{eq_5}
\end{eqnarray}

Provided that $\lambda^2 > \frac{\xi-1}{2}$ and $\xi>0$, $|\Delta(n)| < 1$ and therefore by the binomial theorem
\begin{eqnarray}
\frac{1}{(1+\Delta(n))^s} = \sum_{k=0}^\infty \frac{\Gamma(k+s)}{\Gamma(s) \ k!} \ \left[-\Delta(n)\right]^k \ .
\nonumber
\end{eqnarray}

Using this result in equation~(\ref{eq_4}) we have
\begin{eqnarray}
\overline{\zeta}(u,s,\xi)  &=& \frac{1}{\xi^s} + \sum_{n=1}^\infty \sum_{k=0}^\infty \frac{\Gamma(k+s)}{\Gamma(s) \ k!} 
\ \sum_{j=0}^k \ \left( \begin{array}{c}
k \\
j \\
\end{array} \right) \frac{\lambda^{2 (k-j)}}{(1+\lambda^2)^{s+k}} \frac{(-\xi)^j}{n^{u (s+j)}}  \ .
\label{eq_6}
\end{eqnarray}

As the sum over $n$ and $k$ converges absolutely, we can sum over $n$ and obtain
\begin{eqnarray}
\overline{\zeta}(u,s,\xi)  &=& \frac{1}{\xi^s} + 
\sum_{k=0}^\infty \frac{\Gamma(k+s)}{\Gamma(s) } \sum_{j=0}^k \ \frac{(-\xi)^j}{j! (k-j)!} 
\frac{\lambda^{2 (k-j)}}{(1+\lambda^2)^{s+k}}  \zeta(u (s+j)) \ ,
\nonumber
\end{eqnarray}
which completes our proof.
\end{pf*}

Having proved our fundamental result, equation~(\ref{eq_2}), we now discuss some of the properties of the new
series. We first stress that (\ref{eq_2})  is independent of the arbitrary parameter 
$\lambda$, although $\lambda$ appears explicitly in the expression. This happens because 
equation~(\ref{s3_1a}) is independent of $\lambda$ and it has just been proved that our new series, 
equation~(\ref{eq_2}) converges to equation~(\ref{s3_1a}) provided that $\lambda^2 > (\xi-1)/2$ and $\xi>0$.
In other words we can say that equation~(\ref{eq_2}) describes a family of series, each corresponding to a 
different value of $\lambda$ and each converging to the same series, equation~(\ref{s3_1a}).

Consider the partial sum
\begin{eqnarray}
\overline{\zeta}^{(N)}(\lambda,u,s,\xi)  &=& \frac{1}{\xi^s} + 
\sum_{k=0}^N \frac{\Gamma(k+s)}{\Gamma(s) } \ \Psi_{k}(\lambda,u,s,\xi) \nonumber
\end{eqnarray}
obtained by restricting the infinite sum to the first $N+1$ terms. Obviously 
$\overline{\zeta}^{(N)}(\lambda,u,s,\xi)$ depends upon $\lambda$ as a result of having
neglected an infinite number of terms. We can use this feature to our advantage and fix
$\lambda$ so that the convergence rate of the series is maximal.

The proper value of $\lambda$ is chosen using the PMS, which amounts to find $\lambda$
fulfilling the equation
\begin{equation}
\frac{d}{d\lambda} \ \overline{\zeta}^{(N)}(s,\lambda) = 0 \ .
\end{equation}

A straightforward mathematical interpretation of this condition is that the value of $\lambda$ complying
with this equation also minimizes the difference~\cite{[5]} 
\begin{equation}
\Xi \equiv \left[\overline{\zeta}(u,s,\xi) - \overline{\zeta}^{(N)}(\lambda,u,s,\xi) \right]^2 \ .
\end{equation}

To lowest order, which corresponds to choosing $N=1$, one obtains the optimal value 
\begin{equation}
\lambda_{PMS}^{(1)} = \sqrt{\xi  \ \frac{\zeta( u (1+ s))}{ \zeta(s u)}} ,
\label{eq_pms}
\end{equation}
which  can be used as long as $\lambda_{PMS}^2>(\xi-1)/2$. Notice that, since $\lambda_{PMS}$ depends 
upon $\xi$ then $\overline{\zeta}^{(N)}(\lambda_{PMS},u,s,\xi)$ will not be a polynomial in $\xi$. On the
other hand, if we had chosen $\lambda=0$, then  $\overline{\zeta}^{(N)}(0,u,s,\xi)$ would be a polynomial
of $N^{th}$ order in $\xi$. In that case however the convergence of the series would be strictly limited to
the region $\xi<1$. For this reason we will refer to our accelerated series corresponding to $\lambda_{PMS}$
and to $\lambda=0$ as being ``nonperturbative'' and ``perturbative'' respectively.

\begin{figure}
\begin{center}
\includegraphics[width=10cm]{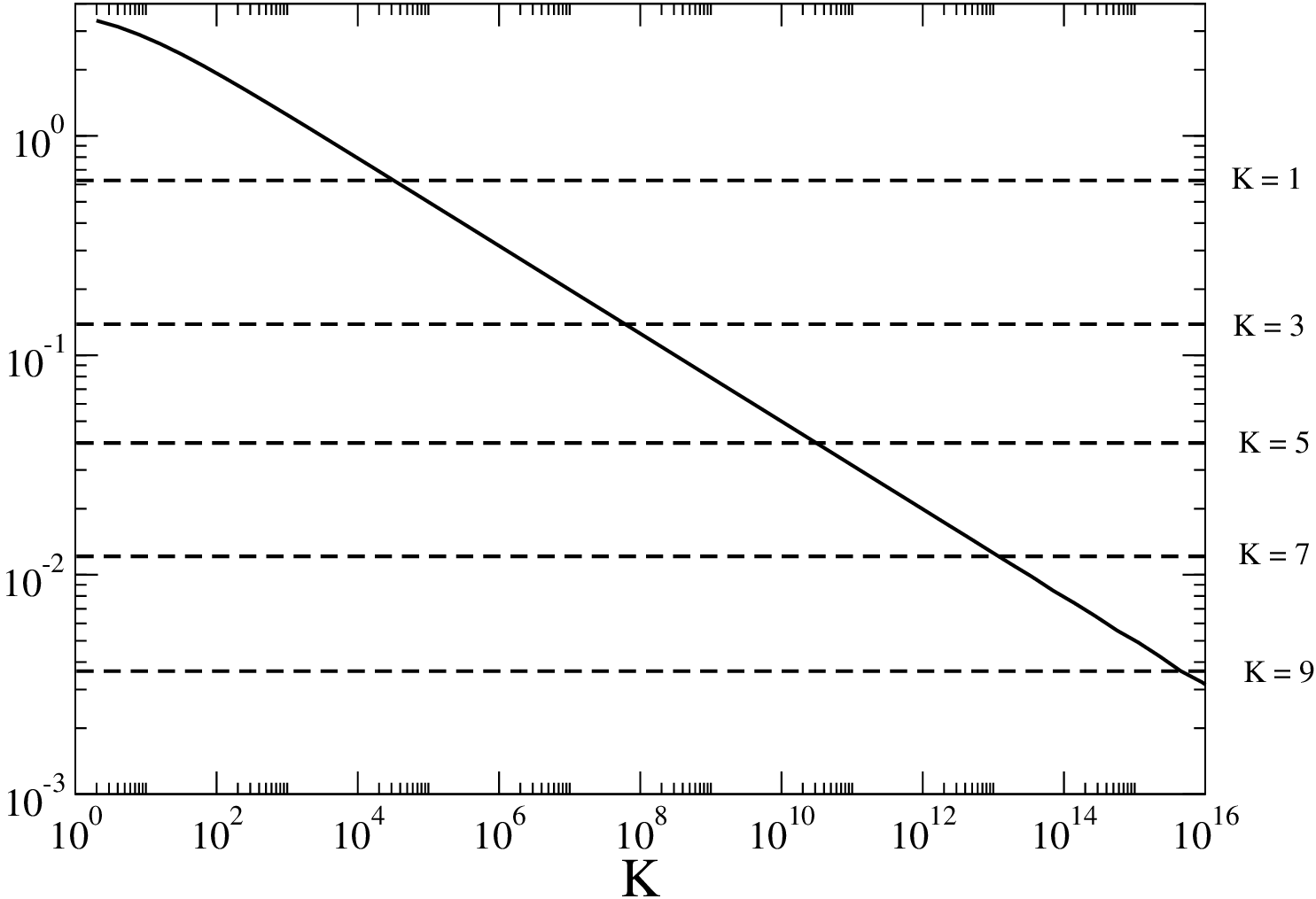}
\caption{Difference  $|\overline{\zeta}^{(K)}(2,\frac{3}{5},1)-\overline{\zeta}(2,\frac{3}{5},1)|$  
as a function of the number of terms in the sum.\label{fig7}}
\end{center}
\end{figure}

In Figure~\ref{fig7} we plot the difference  $|\overline{\zeta}^{(K)}(2,\frac{3}{5},1)-\overline{\zeta}(2,\frac{3}{5},1)|$ 
as a function of the number of terms in the sum of equation (\ref{s3_1a}).  In this case $su=6/5$ is close to $1$ and 
(\ref{s3_1a}) converges very slowly.
The horizontal lines are the values obtained by using equation~(\ref{eq_2}) with the optimal value given in the 
equation (\ref{eq_pms}).

\section{Conclusions}
\label{concl}

The variational approach that we have described in this paper allows in many cases to convert a slowly converging
series into a series which converges exponentially. In most cases the results that are obtained by 
following this approach are explicit. Although we have considered only a few examples, 
we believe that it should be possible to apply this method to a larger class of series. 
Indeed the results of ~\cite{AS:04,AAFS:04}, which focus on the calculation of the period of a classical oscillator,
and of ~\cite{AL:04}, which focusses on the calculation of the spectrum a quantum oscillator in the WKB approximation, 
provide a similar series representation for elliptic functions. In \cite{AM_finite_T} the author has also applied 
the accelerated series for the generalized Hurwitz zeta function obtained in the present paper 
to the calculation of bosonic one loop integrals  at finite temperature in quantum field theory. 
The author and collaborators are presently applying the results
of this paper to the calculation of the Casimir energy for massive scalar fields between parallel plates~\cite{AAH05}.
In our opinion, the strongest results of the present paper are equations (\ref{s4_3})
and (\ref{s4_4}); it remains to study the possible uses of such equations both in mathematical and physical problems
(for example, these equations could be used to further improve the rate of convergence of series such as the one in
equation~(\ref{eq4})).

\end{document}